\documentclass[12pt,a4paper]{article}
\usepackage{amsfonts,latexsym}
\usepackage{amsmath,amssymb}
\usepackage{graphicx,color}

\renewcommand{\thefootnote}{\fnsymbol{footnote}}

\begin{document}

\vspace{12mm}

\begin{center}
{{{\Large {\bf Comment on Quantum Massive Conformal Gravity }}}}\\[10mm]

{Yun Soo Myung\footnote{e-mail address: ysmyung@inje.ac.kr}}\\[8mm]

{Institute of Basic Sciences and Department  of Computer Simulation, Inje University Gimhae 621-749, Korea\\[0pt]}

\end{center}
\vspace{2mm}

\begin{abstract}
In a recent paper in EPJC March 2016, Faria has shown that quantum
massive conformal gravity is renormalizable but has ghost states. We
comment  this paper on the aspect of renormalizability.
\end{abstract}
\vspace{5mm}

\vspace{1.5cm}

\hspace{11.5cm}{Typeset Using \LaTeX}
\newpage
\renewcommand{\thefootnote}{\arabic{footnote}}
\setcounter{footnote}{0}

Recently, a paper in EPJC March 2016~\cite{Faria:2015vea}, Faria has
insisted that quantum massive conformal gravity is renormalizable
but has ghost states.  The proposed action for this purpose  is just
the massive conformal gravity
action(MCG)~\cite{Faria:2013hxa,Myung:2014aia}
\begin{eqnarray}S_{\rm MCG}=-\frac{1}{12 }\int d^4 x\sqrt{-g}
\Big[\Big(\phi^2R+
6\partial_\mu\phi\partial^\mu\phi\Big)-\frac{1}{m^2}C^2\Big],~~C^2=C^{\mu\nu\rho\sigma}C_{\mu\nu\rho\sigma}
\label{MCG}
\end{eqnarray}
which  is classically invariant under the  conformal transformations
as
\begin{equation} \label{cft}
g_{\mu\nu} \to \Omega^2(x)g_{\mu\nu},~~\phi \to \frac{\phi}{\Omega}.
\end{equation}
Here $\Omega(x)$ is an arbitrary function of the spacetime
coordinates. We note here that the MCG action ($S_{\rm MCG}$) is
composed of the conformal dilation gravity ($S_{\rm CDG}$) and the
conformal gravity ($S_{\rm CG}$). Since the integral of the Euler
density
($E=R_{\mu\nu\rho\sigma}R^{\mu\nu\rho\sigma}-4R_{\mu\nu}R^{\mu\nu}+R^2$)
is a topological invariant quantity, $S_{\rm CG}$ reduces to the
Weyl gravity
\begin{equation} \label{CDG}
S_{\rm WG}=\frac{1}{6 m^2}\int d^4x \sqrt{-g}
\Big[R_{\mu\nu}R^{\mu\nu}-\frac{1}{3}R^2\Big]\equiv
\frac{1}{6m^2}\int d^4x \sqrt{-g}\tilde{C}^2.
\end{equation}
It was argued that the massive conformal gravity (\ref{MCG}) is a
renormalizable quantum theory of gravity which has two massive ghost
states. However, Faria's work is far from showing that (\ref{MCG})
is a renormalizable quantum gravity because it  was  based on
performing the canonical quantization of the  second-order bilinear
action. Concerning the renormalizability, Faria has mentioned that
the graviton propagator of
$\Psi_{\mu\nu}=h_{\mu\nu}-\eta_{\mu\nu}h/2$,
\begin{equation} \label{gpro}
D^{\mu\nu,\alpha\beta}_\Psi=-\frac{i}{2}\Big(\eta^{\mu\alpha}\eta^{\nu\beta}
+\eta^{\mu\beta}\eta^{\nu\alpha}-\eta^{\mu\nu}\eta^{\alpha\beta}\Big)
\int
\frac{d^4p}{(2\pi)^4}\frac{m^2e^{-ip\cdot(x-y)}}{(p^2-i\chi)(p^2+m^2-i\chi)}
\end{equation}
has a good behavior of $1/p^4$ at high momenta, making (\ref{MCG})
power-counting renormalizable in the Minkowski spacetime. Unless the
massive pole is shown to be unphysical, the MCG (\ref{MCG}) is
perturbatively meaningless because the graviton propagator
(\ref{gpro}) has ghost state. We wish to point out that this was a
clear observation since the seminal work of Stelle released 40 years
ago~\cite{Stelle:1976gc}.

 It was known that the MCG (\ref{MCG}) is a promising quantum
(gravity) model because the conformal symmetry restricts the number
of counter-terms arising from the perturbative quantization of the
scalar (dilaton) $\phi$, while keeping the graviton
fixed~\cite{'tHooft:1974bx}. The inclusion of  conformal symmetry is
the reason for the cancelations.  Explicitly, the one-loop counter
term of the conformal dilation gravity ($S_{\rm CDG}$) is given
by~\cite{tHooft:2011aa,Alvarez:2014qca}
\begin{equation} \label{ct}
\Gamma_{\rm CDG}=-\frac{1}{960\pi^2(n-4)}\int
d^nx\sqrt{-g}\Big[R_{\mu\nu}R^{\mu\nu}-\frac{1}{3}R^2\Big],
\end{equation}
which is proportional to  the $S_{\rm WG}$ (\ref{CDG}), reflecting
the conformal symmetry. We don't need any counter terms more. In the
absence of conformally coupled term ($\phi^2R$), however, quantizing
a purely kinetic term requires (\ref{ct}) as well as $R^2$-term
(conformally non-invariant term). Hence (\ref{MCG}) plays the role
of a proper quantization action when quantizing the dilation $\phi$
in the fixed curved background $g_{\mu\nu}$.

 On the other hand, Stelle has
proposed  the conformally non-invariant Lagrangian  of $\sqrt{-g}(
R+a \tilde{C}^2+b R^2)$  to improve the perturbative properties of
Einstein gravity~\cite{Stelle:1976gc}. The first two terms could be
derived from (\ref{MCG}) by making use of the conformal
transformation (\ref{cft}).  Also, one includes  $R^2$ term as a
counter term of $g_{\mu\nu}$. If $ab\not=0$, the renormalizability
was achieved but the unitarity was violated, which shows  that the
renormalizability is not compatible with the unitarity. Although the
Weyl-squared term ($\tilde{C}^2$) improves the ultraviolet
divergence, it induces ghost excitations which spoil the unitarity
simultaneously.  Here, the price one has to pay for making the
theory renormalizable is the loss of unitarity.

Up to now, there is no obvious way  to attain the renormalizability
without violating the unitarity in quantizing the gravity.

\end{document}